\documentclass[twocolumn]{aastex631}
\usepackage{txfonts}
\usepackage{graphicx}
\graphicspath{{./}{figures/}}
\usepackage{natbib}

\newcommand\aastex{AAS\TeX}

\received{}
\accepted{}
\submitjournal{ApJ}

\shorttitle{\aastex Magnetic activity in the host star GJ 504}
\shortauthors{Di Mauro et al.}
\begin{document}

%\title{On the non-detection of solar-like pulsations in the host star GJ~504 observed by TESS}
\title{On the characterization of GJ~504: a magnetically active planet-host star observed by\\
the Transiting Exoplanet Survey Satellite (TESS)}

\correspondingauthor{Maria Pia Di Mauro}
\email{maria.dimauro@inaf.it}

\author[0000-0001-7801-7484]{Maria Pia Di Mauro }
\affil{INAF-IAPS, Istituto di Astrofisica e Planetologia Spaziali
Via del Fosso del Cavaliere 100 
00133 Roma, Italy}

\author[0000-0001-8623-5318]{Raffaele Reda}
\affil{Dipartimento di Fisica, Università di Roma Tor Vergata, Via della Ricerca Scientifica 1, 00133 Roma, Italy}
\affil{INAF-IAPS, Istituto di Astrofisica e Planetologia Spaziali
Via del Fosso del Cavaliere 100 
00133 Roma, Italy}

\author[0000-0002-0129-0316]{Savita Mathur}
\affiliation{Instituto de Astrof\'isica de Canarias (IAC), 38205 La Laguna, Tenerife, Spain}
\affiliation{Universidad de La Laguna (ULL), Departamento de Astrof\'isica, E-38206 La Laguna, Tenerife, Spain}

\author[0000-0002-8854-3776]{Rafael ~A.~Garc\'\i a}
\affiliation{AIM, CEA, CNRS, Universit\'e Paris-Saclay, Universit\'e de Paris, Sorbonne Paris Cit\'e, F-91191 Gif-sur-Yvette, France}
%\affiliation{AIM, CEA, CNRS, Universit\'e Paris-Saclay, Universit\'e Paris Diderot, Sorbonne Paris Cit\'e, F-91191 Gif-sur-Yvette, France}

\author[0000-0002-1988-143X]{Derek L. Buzasi}
\affiliation{Dept. of Chemistry \& Physics, Florida Gulf Coast University, 10501 FGCU Blvd. S., Fort Myers, FL 33965 USA}

\author[0000-0001-8835-2075]{Enrico Corsaro}
\affil{INAF-Astrophysical Observatory of Catania,
Via S. Sofia 78
95123 Catania, Italy}

\author[0000-0001-9405-5552]{Othman Benomar} 
\affiliation{National Astronomical Observatory of Japan, Mitaka, Tokyo, Japan}
\affiliation{Center for Space Science, New York University Abu Dhabi, UAE}

\author[0000-0002-1241-5508]{Luc\'\i a Gonz\'alez Cuesta}
\affiliation{Instituto de Astrof\'isica de Canarias (IAC), 38205 La Laguna, Tenerife, Spain}
\affiliation{Universidad de La Laguna (ULL), Departamento de Astrof\'isica, E-38206 La Laguna, Tenerife, Spain}

\author[0000-0002-3481-9052]{Keivan G. Stassun} 
\affiliation{Vanderbilt University, Department of Physics \& Astronomy, 6301 Stevenson Center Ln., Nashville, TN 37235, USA}
\affiliation{Vanderbilt Initiative in Data-intensive Astrophysics (VIDA), 6301 Stevenson Center Lane, Nashville, TN 37235, USA}

\author[0000-0002-4638-3495]{Serena Benatti}
\affil{INAF - Osservatorio Astronomico di Palermo, Piazza del Parlamento 1, 90134 Palermo, Italy}
\author[0000-0002-2662-3762]{Valentina D'Orazi}
\affil{INAF Osservatorio Astronomico di Padova, vicolo dell’Osservatorio 5
35122, Padova
Italy}
\affil{School of Physics and Astronomy, Monash University, Clayton, VIC 3800, Australia}

\author[0000-0001-7369-8516]{Luca Giovannelli}
\affil{Dipartimento di Fisica, Università di Roma Tor Vergata, Via della Ricerca Scientifica 1, 00133 Roma, Italy}
\affil{INAF-IAPS, Istituto di Astrofisica e Planetologia Spaziali
Via del Fosso del Cavaliere 100
00133 Roma, Italy}

\author[0000-0001-8467-1933]{Dino Mesa}
\affil{INAF Osservatorio Astronomico di Padova, vicolo dell’Osservatorio 5
35122, Padova
Italy}

\author[0000-0002-7399-0231]{Nicolas Nardetto}
\affil{Universit\'e C\^ote d'Azur, Observatoire de la C\^ote d'Azur, CNRS, Laboratoire Lagrange, France}



\begin{abstract}
We present the results of the analysis of the
photometric data collected in long- and short-cadence mode by the Transiting Exoplanet Survey Satellite (TESS) for GJ\,504,  a well-studied planet-hosting solar-like star, whose fundamental parameters have been largely debated during the last decade.
Several attempts have been made by the present authors to isolate the oscillatory properties expected on this main-sequence star, but
%with mass $M=(1.28\pm0.07)\mathrm{M}_{\odot}$, radius $R= (1.38\pm0.2)\mathrm{R}_{\odot}$ and age  $\leq  2.6$~Gyr, as predicted by theoretical models. 
%we found only a  marginal hint of the presence of pulsations around $\simeq 2000~ \mu$Hz  {\bf with no statistical significance.}
%a large separation $\Delta \nu= (85.0\pm3.6) \mu$Hz.
 we did not find any presence of solar-like pulsations.
The suppression of the amplitude of the acoustic modes can be explained by 
the high level of magnetic activity revealed for this target, not only by the study of the photometric light curve but also by the analysis of three decades of available of Mount Wilson spectroscopic data.
In particular, our measurements 
of the stellar rotational period  $P_{\mathrm rot}\simeq 3.4$~days and 
of the main principal magnetic cycle of $\simeq 12$~a confirm previous findings and
allow us to locate this star in the early main-sequence phase of its evolution during which the
chromospheric activity  is dominated by the superposition of several cycles  before the transition to the phase of the magnetic-braking shutdown with the subsequent decrease of the magnetic activity.
\end{abstract}


\section{Introduction}

Over the last decade, thanks to the successful photometric space missions Convection, Rotation, and Transits \citep[CoRoT;][]{Baglin06} and Kepler/K2 \citep{Boroucki10}  mainly conceived for exoplanets, but extremely suitable for detection of stellar pulsations, asteroseismology 
has produced an extraordinary revolution in astrophysics \citep[e.g.][]{Becketal2012, Bedding2011, Silvaguirre, Stello2016}. This unveiled a wealth of results on the physical properties of stars over a large part of the H-R 
diagram and mostly for solar-like stars, which exhibit  pulsations excited by near-surface turbulent convection, as it happens in the Sun.

The extreme photometric precision made these missions spectacularly successful also in their primary goal: the detection and characterization of extrasolar planetary systems by using the transit technique \citep{Borucki2013}. 
 Thus, in recent years, a flood of very high-quality data has been collected, and the search for new worlds is in progress, and we are living exciting times in this respect. 
Despite the incredible effort in refining observational and post-processing techniques, our interpretation and comprehension of planetary systems architecture, formation, and evolution mechanisms heavily relies on the accuracy of the inferred characteristics of the host stars and the effects on their planets \citep[]{vaneylen2014, Borucki2013, Huber2013, Chaplin2013}. This often represents a significant challenge, especially for isolated field stars \citep{Soderblom}.

The most recently launched NASA space mission Transiting Exoplanet Survey Satellite \citep[TESS;][]{Ricker14},  is poised to continue the synergy between asteroseismology and exoplanet science, enlarging the held of asteroseismic inference to full-sky. Indeed, with the original high-cadence mode of 120 s (Nyquist frequency of 4166 $\mu$Hz) used during the first two years of its main mission and the newer fast cadence of 20 s that started during the extended mission, TESS should be able to detect oscillations in many main-sequence solar-like stars in spite of their low intrinsic amplitudes of parts-per million \citep[ppm;][]{Garcia2019}.

According to the TESS Asteroseismic Target List \citep{Campante16, Schofield19}, thousands of main-sequence and subgiant solar-like stars should show detectable modes. So far,  signatures of such oscillations have been detected only in a handful of solar-like stars \citep[e.g.,][]{gandolfi2018, huber2019, chontos2020, metcalfe2020, addison2021, metcalfe2021}. 
Recently, \citet{huber2021} compared the power spectra of three stars observed by TESS with both cadences of 2 minutes and 20 s. While the modes were barely visible with the 2 minutes cadence, the faster cadence drastically increased by $ \sim  30\%$ the signal-to-noise ratio allowing the characterization of the individual modes. Part of this improvement could be explained by the difference of the cosmic-ray rejection applied to both cadences. However, the difficulties to detect the oscillation modes could also be due to the properties of the stars.

The nondetection of modes has been investigated also in many stars observed by the \emph{Kepler} mission. For solar-like stars, the usual explanation is the surface magnetic activity of the star \citep[e.g.,][]{Chaplin2011}
as it is known that a high level of magnetic activity can reduce the amplitude of the modes \citep{Garcia2010, Kiefer2017,Santos2018}.
Nevertheless this is not the only culprit as shown in \citet{Mathur2019}. Indeed metallicity or binarity can also have an impact on the amplitude of the modes \citep{Gaulme2020}. For the binarity explanation, \citet{Gaulme2020} and \citet{Benbakoura2021} showed that synchronized binaries can have enhanced magnetic activity, which subsequently leads to suppressed modes. So in that case, magnetic activity is again the origin of the nondetection.

Here, we present the result of the analysis of 
the solar-type star GJ 504 (spectral type G0), which was observed by the TESS mission for 27 days during sector 23 from 2020 March 18 to 2020 April 16, 2020, 2 minutes cadence mode and again during sector 50 for 27 days from 2022 March 26 to 2022 April 22, 2022, using 20 s cadence mode.
Furthermore, spectroscopic observations from the Mount Wilson Observatory (MWO) are also studied to better characterize the surface magnetic activity of the star.

 GJ~504 is considered a very interesting case study,  claimed to host a substellar companion whose nature is strongly debated. Nevertheless, asteroseismology might provide the only powerful mean to dissolve any doubts about the evolutionary state of this target and hence on the identity of the secondary object. In fact, the detection of typical signatures of solar-like oscillations in the power spectrum would define with good accuracy the age and all the physical parameters of this low-mass star \citep[e.g.,][]{dimauro2004,dimauro16}. There are many methods
to estimate the age of a single star \citep{Soderblom}: empirical indicators such as stellar activity and gyrochronology, which link rotation to age \citep[e.g.,][]{Skumanich72, Barnes2007, Mamajek08}; photospheric lithium abundance \citep[e.g.][]{Li2012}; comparison of stellar model
isochrones with observed classical parameters \citep[e.g.][]{Pont2004}. However, the accuracy that can currently be reached by using all these methods is not satisfactory, not only because of the large errors in the estimates
but also because better precision and accuracy can be reached only by using seismic diagnostics \citep[see, e.g.,][]{Metcalfe2010, Lebreton2014, lebretongu14}.

This paper is organized in the following sections:
Section 2 introduces the reader to the target presenting the spectroscopic fundamental parameters and the theoretical predictions deduced by means of stellar evolutionary models and asteroseismic scaling laws; Section 3 presents the observations and the data calibration used in this work; in Section 4, we study the surface rotation and magnetic activity of the this star; in Section 5, we describe the search for solar-like oscillations; Section 6 discusses the reasons for the nondetection of solar-like oscillations; Section 7 shows the conclusions on our attempt to characterize the structure of this star.

\section{The solar-like star GJ~504}
\subsection{An intriguing case}
During the last 25 yr, several dedicated space missions, together with great developments in observational techniques, have allowed huge progresses in the search for new worlds outside the solar system. In particular, besides statistics,  several hundreds of bright stars have been monitored and multi-wavelengths data collected to understand and characterize the formation and the evolution of the already-discovered planetary systems.

A controversial case still debated today is represented by the solar-type star GJ~504 (HD 115383, TIC 397587084), a G0-type star with $T_{\mathrm{eff}}\approx6200\;K$, which appears to be a little more massive than the Sun \citep{Kuzuhara13, DOrazi17}, with a rotational period $P_{\mathrm{ rot}}=3.329$\,days, average of the values reported by \citet{Messina03} and \citet{Donahue96}.
In 2013, by exploiting high-contrast near-IR and L'-band observations as part of the SEEDS survey, \citet{Kuzuhara13} reported the direct-imaging discovery of a Jovian planet orbiting this star, with a projected separation of 43.5 AU. Unfortunately, no radial velocity data are available for the host star, and hence the mass of the planet needs to be estimated in other ways.
Employing the gyrochronology technique, based on the stellar chromospheric activity indices (as given by the Ca II H and K emission lines) and on X-ray observations (the star is included in the ROSAT catalog), \citet{Kuzuhara13} estimated the age of GJ~504 to be $Age=160^{+350}_{-60}$\,Myr. Under this age assumption, the comparison of the observed planet's luminosity at each band with the theoretical models by \citet{2003Baraffe} implies that the mass of the substellar companion  (named GJ~504b) should be $M_{\rm P}=4^{+4.5}_{-1.0}\,M_\mathrm{Jup}$. According to \citet{Kuzuhara13}, the measured characteristics make GJ~504b a very interesting object because it represents the first example of giant planet on a wide orbit around a solar-type star. Moreover, the planet appears to be significantly cool ($T_\mathrm{eq} = 510^{+30}_{-20}\,\mathrm{K}$), with an almost-cloud-free atmosphere due to its blue color (J-H=-0.23) and as reported by \citet{Janson13}, it represents the first known extrasolar planet with methane-dominated atmosphere (T-type).

However, afew years later the young age of GJ~504 has been disproved by \citet{Fuhrmann15}, thanks to evidences arising from high-resolution and high-quality spectra. In fact, the authors derived a stellar gravity of $\log g=4.23$ dex (from the Hipparcos parallax and adopting spectroscopic temperature), which results to be not compatible with a stellar age of few hundreds Myr. This gravity estimate is also in agreement with the value of $\log g=4.17 $ dex previously determined in \citet{Fuhrmann04}, based on the spectral fitting of Mg Ib lines and several independent spectroscopic studies and position of the star in the color-magnitude diagram \citep[see, e.g.,][]{dasilva2012}. This picture implies that GJ~504 should be a star with approximately the solar age, and as a consequence, the companion has to be identified as a brown dwarf ($M_{\rm P} \sim 25 \mathrm{M}_\mathrm{Jup}$) rather than a giant planet. To explain the relatively high stellar rotation velocity and chromospheric activity level of the host star, and reconcile the isochronal ages with direct indicators, \citet{Fuhrmann15} invoked a merging event. GJ~504 might have engulfed a substellar companion that is responsible for speeding up the rotational velocity and accounts for the enhanced activity levels \citep{Oetjens2020}. 

\citet{DOrazi17}, reassessing the properties of 
GJ~504, have found that the surface gravity of the star implies an evolutionary stage obtained by the isochrones comparison, which suggests an age range between 1.8 and 3.5 Gyr (most probable age $\approx$ 2.5 Gyr). To reconcile all the age indicators and to explain the high level of activity, also these authors suggest a merging scenario (more recent than 200 Myr) with a very close hot Jupiter companion. 

The system has been recently revisited by \citet{Bonnefoy18} by using interferometric, radial-velocity, and high-contrast imaging observations. They found an interferometric radius of $R=(1.35 \pm 0.04)\mathrm{R}_{\odot}$ for GJ~504, which is compatible with two isochronal age ranges $(21 \pm 2)$~Myr and $(4.0 \pm 1.8)$~Gyr. According to this work, the mass of GJ~504b is expected to be $M_{\rm P} = 1.3^{+0.6}_{-0.3}\,M_\mathrm{Jup}$ for the young-age case and $M_{\rm P} = 23^{+10}_{-9}\,M_\mathrm{Jup}$ for the old one.

Therefore, the evolutionary stage and the age of 
GJ~504 is still an open problem with no clear solution to date. This uncertainty closely concerns the mass estimation of the star's companion, which could be a Jovian planet or a brown dwarf. In addition, \citet{Skemer16}, through a photometric study, suggested for GJ~504b a higher metallicity ($\mathrm{[M/H]}_{p}\approx +0.6$) with respect to the host star ($\mathrm{[M/H]}_{\star}\approx+0.1-0.3$), adding to this system another element of interest. 

For all the above mentioned reasons, the GJ~504 system constitutes an intriguing case, which deserves to be carefully studied, with the aim of shedding light on the age of the star and, accordingly, on the nature of the substellar companion.

\subsection{Fundamental parameters}
\label{sec2.2}
\begin{figure}[ht!]
    \centering
   \includegraphics[width=6.5cm, angle=90]{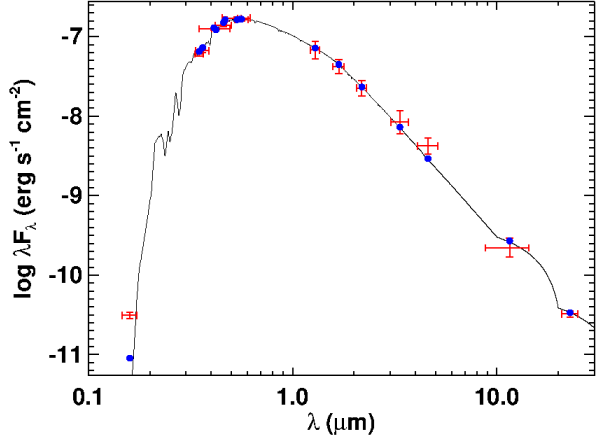}
    \caption{Spectral energy distribution. Red symbols represent the observed photometric measurements, where the horizontal bars represent the effective width of the pass-band. Blue symbols are the model fluxes from the best-fit Kurucz atmosphere model (black). 
\label{fig:sed}}
\end{figure}

With the aim to properly characterize this star,
we performed an analysis of the broadband spectral energy distribution (SED) together with the Gaia EDR3 \citep{Gaia2018} parallax measurement
following the procedures described in \citet{Stassun16}, \citet{Stassun17}, and \citet{Stassun18}.
 The input parameters and the obtained results are summarized in Table \ref{tab1}.   We employed the $UBV$ magnitudes from \citet{Mermilliod06}, the $B_T V_T$ magnitudes from {\it Tycho-2}, the Str\"omgren $uvby$ magnitudes from \citet{Paunzen15}, the $JHK_S$ magnitudes from {\it 2MASS}, the W1--W4 magnitudes from {\it WISE}, and the FUV magnitude from {\it GALEX}. The available photometry, all together, spans the full stellar SED over the wavelength range 0.2\,--\,22~$\mu$m (Fig.~\ref{fig:sed}). 

\begin{table}

\begin{center}
\caption{The fundamental parameters of GJ~504 (TIC 397587084).\label{tab1}}
\renewcommand{\tabcolsep}{0mm}
\begin{tabular}{l c}
\tableline\tableline
\noalign{\smallskip}
\multicolumn{2}{c}{Basic Properties} \\
\noalign{\smallskip}
\hline
\noalign{\smallskip}
%Hipparcos ID &   GJ 504\\
%TIC ID & 397587084 \\
%Right Ascension	& 19 37 25.575 \\
%Declination	& +38 56 50.515 \\
%$V$ Magnitude & 5.21\pm0.05 \\
TESS  Magnitude & $4.6552 \pm 0.0073^{a}$  \\
%$K$ Magnitude &  \\
$\pi$ (mas) &  $57.0186 \pm 0.2524^{b}$ \\
\noalign{\smallskip}
\hline
\noalign{\smallskip}

\multicolumn{2}{c}{Spectroscopic parameters$^{c}$} \\
\noalign{\smallskip}
\hline
\noalign{\smallskip}
$T_{\mathrm{eff}}$  (K) & $6205\pm 20 $ \\
$[\mathrm{ Fe/H}]$ (dex) & $0.22\pm0.04$\\
$v \sin i$ (km\,s$^{-1}$) & $2.8 \pm 1.6$ \\
 $\log g$ (dex) & $4.29\pm 0.07$ \\
\noalign{\smallskip}\hline
\hline
\multicolumn{2}{c}{SED results} \\
\noalign{\smallskip}
\hline
\noalign{\smallskip}
 
%Bolometric Flux, $
$F_{\rm bol} \, {\mathrm{\,(erg\, s^{-1} cm^{-2})}}$\,\,\, &
\,\,\,
$(2.10\pm 0.02)\cdot 10^{-8} $  \\
 
 $L_{\mathrm SED} /\mathrm{L}_{\sun}$ & $2.01\pm 0.03$ \\
\noalign{\smallskip}

$M_{\mathrm SED} /\mathrm{M}_\sun$& $1.07\pm0.17$  \\
  $R_{\mathrm SED} /\mathrm{R}_\sun$& $1.227\pm 0.012$\\
%Stellar Density, $\rho_\star$ (gcc)& denstar \\
$P_{\mathrm SED}$ (d)& $2.4\pm 1.3$\\
$Age_{\mathrm SED}$ (Gyr) &  $0.2\pm0.2$\\
\noalign{\smallskip}
\hline
\hline
\multicolumn{2}{c}{SBCR results} \\
\noalign{\smallskip}
\hline
\noalign{\smallskip}
 $\theta_LD$ (mas)&$ 0.622 \pm 0.122 $\\
  $R_\mathrm{SBCR} /\mathrm{R}_\sun$& $1.176\pm 0.192$\\
\noalign{\smallskip}
\hline

\end{tabular}
\end{center}
 Notes:  \\ 
$^{a}$ Adopted from the TESS Input Catalog \citep{stassun2019}.\\
$^{b}${Gaia measurement\citep[see][]{Gaia2018}}\\
$^{c}$ {Determined by spectroscopic observations 
\citep[see][]{DOrazi17}}
%$^{b}$ {Determined by observations
%\citep[see][]{Mamajek08}}\\

\end{table}

We performed a fit using Kurucz stellar atmosphere models, with the $T_{\rm eff}$, $\log g$, [Fe/H], and $v\sin i$ taken from the spectroscopic analysis of \citet{DOrazi17}. The remaining parameter is the extinction ($A_V$), which we fixed to be zero due to the star's proximity. The resulting fit is shown in Fig.~\ref{fig:sed}, obtained with a reduced $\chi^2= 1.9$. Integrating the SED fitting model, we obtain the bolometric flux at Earth of \mbox{$F_{\rm bol} =  (2.096 \pm 0.024) \cdot 10^{-7}$ erg~s$^{-1}$~cm$^{-2}$}. Taking the $F_{\rm bol}$ and $T_{\rm eff}$ together with the {\it Gaia\/} parallax, with no adjustment for systematic parallax offset \citep[see, e.g.,][]{Stassun21}, gives the stellar radius as $R_{\rm SED} = (1.227 \pm 0.012)\mathrm{R}_\odot$. The $F_{\rm bol}$ and parallax also yield directly the bolometric luminosity, $L_{\rm SED} = (2.01 \pm 0.03)\mathrm{L}_\odot$. 

The empirical stellar radius determined above affords an opportunity to estimate the stellar mass empirically as well, via the spectroscopically determined surface gravity, obtaining $M_{\rm SED} = (1.07\,\pm\,0.17)$~M$_\odot$. This value is consistent with that estimated via the eclipsing-binary based relations of \citet{Torres10}.
%, which gives 
%\mbox{$M_\star = %(1.25\,\pm\,0.08)$~M$_\odot$}. 

Using the activity-age relations of \citet{Mamajek08}, we obtained from $R'_{\rm HK}$ and the star's $B-V$ color, an age of $Age_{\star} = (0.2 \pm 0.2)$~Gyr and a rotational period for the star of \mbox{$P_{\mathrm SED} = (2.4\,\pm\,1.3)$~days} which is consistent with previous findings of 3.3~days \citep{Donahue96, Messina03,Wright11}. 

To get another independent measurement of the stellar radius, it is also possible to employ the approach based on the
Surface-Brightness Colour relationships (SBCR), which allows to easily estimate the limb-darkened angular diameter of the star. The latter combined with the distance of the star provides the linear stellar radius. 

Considering the SBCR from \citet{2021A&A...652A..26S} obtained for late-type dwarf stars (their Table~4), $m_G=5.0398 \pm 0.0029$ mag \citep{2020yCat.1350....0G}, $m_{K_s}=4.033 \pm 0.238$ mag \citep{2003yCat.2246....0C}, and the extinctions in the visual and Gaia bands $A_v=A_g=0.0$~mag derived from \texttt{Stilism} tool \citep{Lallement2014, Capitanio2017}, as well as $A_k=0.089 A_v$ \citep{2009ApJ...696.1407N}, we find $ \theta_{LD}=0.622 \pm 0.014 \pm 0.008 \pm 0.100 $ mas. The uncertainties correspond respectively to the RMS of the SBCR, the uncertainty on the coefficients of the SBCR and the uncertainty on the G and Ks photometries. Using Gaia DR2 parallax, i.e. $\pi=56.8577 \pm 0.1224$ mas \citep{2020yCat.1350....0G}, we obtain $R_{SBCR}=(1.176\pm 0.027 \pm 0.015 \pm 0.19 \pm 0.003)\mathrm{R_{\odot}} =(1.176\pm 0.192)\mathrm{R_{\odot}}$, where  the error $0.003$ is rising from  the uncertainty on the Gaia parallax. This value agrees within the uncertainty with the SED estimate found above.
\subsection{The seismic properties and the asteroseismic prediction by scaling laws}
\label{Sect:pred}

The properties of a solar-like pulsating star
can be described by adopting the
asymptotic development by \cite{tassoul1980}, which predicts that
the oscillations excited in main-sequence stars are
acoustic modes (p modes) with frequencies $\nu_{n,l}$ characterized
by radial order $n$ and harmonic degree $l$, which for 
$l \leq n$ should satisfy the following approximation:
\begin{equation}
\nu_{n,l}\sim\Delta\nu\left(n+\frac{l}{2}+\epsilon \right),
\end{equation}   
where $\epsilon$ is a function of frequency and depends on the properties of the surface layers and $\Delta\nu$, known as the large frequency separation, is the inverse of the sound travel time across the stellar diameter:
%***********************
\begin{equation}
\Delta\nu={\left(2\int_{0}^{R}\!\frac{{\rm d}r}{c}\right)}^{-1},
\label{large}
\end{equation}
%***********************
where $c$ is the local speed of sound at radius $r$ and $R$ is the photospheric stellar radius.
Hence, according to the theory, the solar-like oscillations spectrum of GJ~504 should show a
series of equally spaced peaks separated by $\Delta \nu$  between p
modes of same degree $l$ and adjacent $n$:
\begin{equation}
\Delta\nu \simeq\nu_{n+1,l}-\nu_{n,l}\equiv\Delta\nu_{l}\,.
\label{large}
\end{equation}
In addition, the power spectra of this target should show
 another series of peaks, whose separation
$\delta\nu_{l}$ is
known as the small separation:
\begin{equation}
\delta\nu_{l}\equiv \nu_{n,l}-\nu_{n-1,l+2}
\label{small}
\end{equation}
which is sensitive to the chemical composition gradient in
the central regions of the star and hence to its evolutionary
state.
%*************************
Thus, the determination of the large and small frequency separations from the observed oscillation spectrum can directly provide asteroseismic inferences on the mass
and the age of GJ~504 \citep{CD1988}.  

The observed oscillation power spectrum of the solar-like stars is characterized by a typical Gaussian like
envelope and the frequency of maximum oscillation power is usually indicated by $\nu_{\mathrm max}$.
As conjuctered by \cite{Brown91}, the frequency $\nu_{\mathrm max}$ can be related to the acoustic cutoff frequency $\nu_{ac}$, which defines the upper boundary of the p mode resonant cavities:
\begin{equation}
\nu_{\mathrm max}\propto \nu_{ac}\propto g T_\mathrm{ eff}^{-1/2},
\label{numax}
\end{equation}
Thus, according to Eq. \ref{numax}, the frequency $\nu_{\mathrm max}$  carries information on the physical conditions in the near-surface layers of the star. Thus, as it has been well demonstrated both theoretically \citep{Chaplin2008, Belkacem2011} than observationally \citep{Bedding2003, Stello2008, Bedding2014}, as a solar-type star evolves, its oscillation spectrum moves towards lower frequencies due to the decrease of the surface gravity.
 
To extract a rough estimate of the asteroseismic parameters of the
star to be adopted as guess values for the oscillation analysis, it is possible to assume
well proved scaling-laws as those provided by \citet[see, e.g.,][]{Brown91, Kjeldsen95}, and by \citet{Huber11}. 
These relations, which have been typically calibrated on large samples of main-sequence stars, offer the possibility to  predict  the range of frequencies where the excess of power for a given solar-like star will manifest.
By assuming the relations by \citet{Kjeldsen95} and \citet{Kjeldsen2008}, we calculated  
the value of the expected maximum amplitude of oscillation $A_{\rm max}$ and 
the frequency at the maximum amplitude $\nu_{\rm max}$, using the observed surface gravity $g$ and the effective temperature $T_\mathrm{eff}$ of the star.
By using scaling relations and corrections by \cite{Campante16} we obtained a value for the expected maximum amplitude in the range $A_{\rm max}=(2.51-2.95)$~ppm depending on which input spectroscopic parameters are assumed. 

\begin{deluxetable*}{cccr}
\tablecaption{Predictions for the frequency of maximum oscillation computed assuming spectroscopic data obtained by different authors \label{prediction table}.}
\tablewidth{0pt}
\tablehead{
\colhead{$T_\mathrm{ eff}$} & \colhead{$\log\,g$} & \colhead{$\nu_\mathrm{max}$} & \colhead{Reference} \\
\colhead{($K$)} & \colhead{($\mathrm{cm/s^{2}}$)} & \colhead{($\mu$Hz)} & \colhead{}
}
\startdata
6205 $\pm$ 20 & 4.29 $\pm$ 0.07 & 2096 $\pm$ 338 & \citet{DOrazi17} \\
5978 $\pm$ 60 & 4.23 $\pm$ 0.10 & 1860 $\pm$ 428 & \citet{Fuhrmann15} \\
6130 $\pm$ 48 & 4.33 $\pm$ 0.10 & 2312 $\pm$ 533 & \citet{Maldonado15} \\
6185 $\pm$ 51 & 4.30 $\pm$ 0.07 & 2148 $\pm$ 346 & \citet{Battistini15} \\
5995 $\pm$ 41 & 4.24 $\pm$ 0.02 & 1900 $\pm$ 88 & \citet{Ramirez13} \\
6012 $\pm$ 100 & 4.30 $\pm$ 0.20 & 2179 $\pm$ 100 & \citet{Mishenina13} \\
6234 $\pm$ 25 & 4.60 $\pm$ 0.02 & 4269 $\pm$ 197 & \citet{Valenti05}
\enddata
%\tablecomments{}
\end{deluxetable*}
Table \ref{prediction table} shows the results for the expected $\nu_{\rm max}$ computed assuming spectroscopic measurements published by different authors. Except for the $\nu_{\rm max}$ from the stellar parameters of \citet{Valenti05}, who reported a high surface gravity, all the expected values for the frequency of maximum oscillation lie in the range (1800-2300) $\mu$Hz. Thus, if there is an excess of power due to oscillations in the GJ~504 spectrum, we expect to find it in this range of frequencies.

\subsection{Theoretical prediction by evolutionary models}
\label{models}
Given the observed
fundamental parameters collected in Table \ref{tab1}, it is also possible to face the theoretical challenge
to infer the structural properties  of GJ~504 and predict its detailed oscillation spectrum by constructing stellar evolutionary models which
satisfy the observational constraints.

We produced theoretical structure models for the star by using the ASTEC evolutionary
code \citep{CD08a} by varying the mass and the composition so to match the atmospheric parameters available. 
The resulting evolutionary tracks characterized by fixed
 mass $M$ and
initial chemical composition have been calculated with the OPAL 2005 equation of state 
\citep{OPAL}, OPAL opacities % (Iglesias \& Rogers 1996),
\citep{Igl96}, and the NACRE nuclear reaction rates \citep{NACRE}. 
Convection was treated according to the
mixing-length formalism (MLT) \citep{bohm} and defined through the parameter $\alpha=\ell/H_p$, where $H_p$ is the pressure scale height and $\alpha$ is assumed to be $1.8$. 
The initial heavy-element mass fraction $Z$ in respect to the abundance of the hydrogen $X$
has been calculated from the iron abundance given in Table \ref{tab1} using the relation [Fe/H]$=\log(Z/X)_s-\log({\rm Z/X})_{\odot}$, where $(Z/X)_s$ is the value at
the stellar surface and the solar value was taken to be $(Z/X)_{\odot}=0.0245$ \citep{grevesse}.

\begin{figure}
\centering
\includegraphics[width=8.5cm]{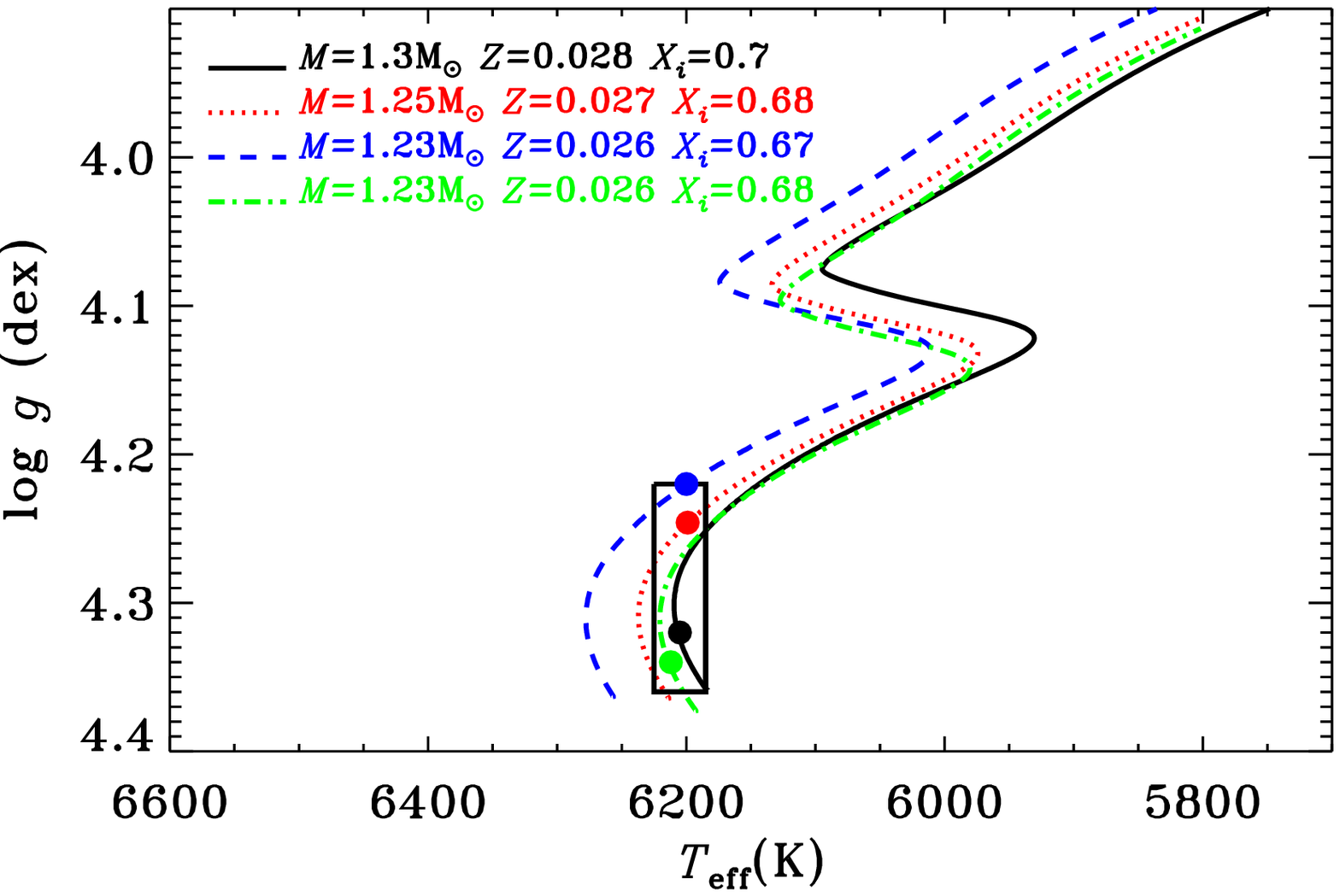}
\includegraphics[width=8.5cm]{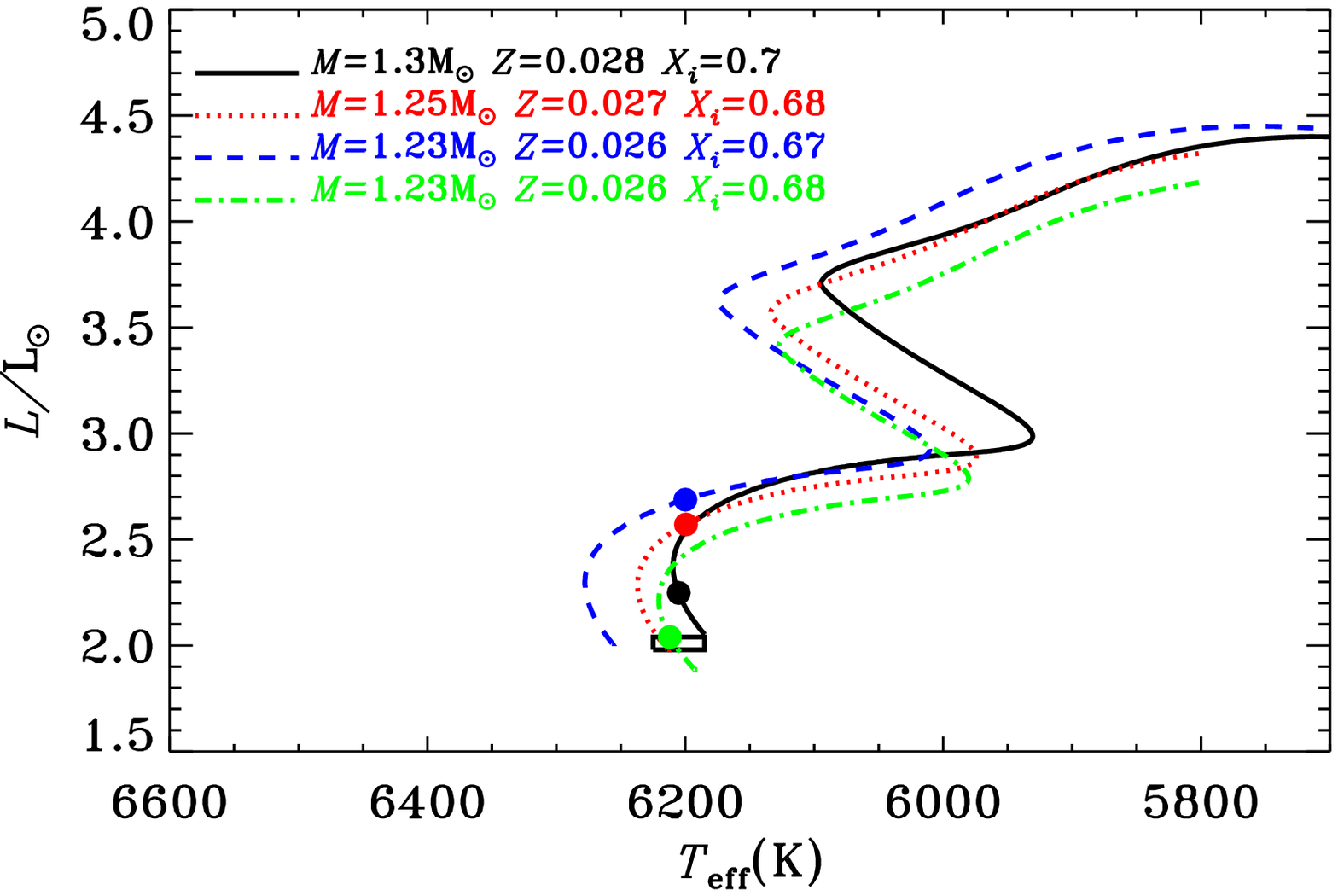}
\caption{Evolutionary tracks plotted in two plans of the H-R diagram calculated for different values of the mass and the metallicity, while all the other parameters are fixed. The initial hydrogen abundance is $X_i$ and the mixing-length coefficient is $\alpha=1.8$. The rectangle defines the one-sigma error box for the observed gravity and effective temperature. 
Coloured dots indicate the position of the four structure models (see Table \ref{tab:fit}) which best reproduce the observations of GJ~504.}
\label{HR}
\end{figure}

Fig.~\ref{HR} shows a series of evolutionary tracks
 obtained for different masses and fixed initial composition, 
plotted in two H-R diagrams, representing respectively the effective temperature-gravity plane and the effective temperature-luminosity plane.
The present evolutionary models do not include additional effects such as overshooting, settling of heavy elements and rotation.

The location of the star in the H-R diagram identifies GJ~504 as being at the beginning of the main sequence phase. In fact, only a small percentage of the hydrogen fuel, indicated by $X_c$ in Table \ref{tab:fit}, has been already converted into helium.
The uncertainty in the observed value of $Z_s$ introduces an uncertainty in the determination of the stellar mass whose value,
considering only the observed spectroscopic parameters, seems to
be limited to the range $ M=(1.28\pm0.07){\rm M_{\odot}}$ hence more massive than the Sun, in agreement with the value predicted by SED analysis (see Section~2.2).
The stellar radius appears $R= (1.38\pm0.20){\rm R_{\odot}}$, a value which is in good agreement within the errors, 
not only with the estimates found in Sec.~\ref{sec2.2} by the SED and the SBCR methods, but also
with the most accurate interferometric radius measured by \citet{Bonnefoy18}.

The age of this star, as obtained from the evolutionary models, can be estimated in the range $0.0-2.6$~Gyr, hence younger than the Sun, so that the convective envelope should appear still quite shallow with a depth not larger than $D_{cz}\simeq0.16 R$. 
Thus, we confirm that GJ~504 is a very young star as we found in Sec. \ref{sec2.2} by SED calculations and
in agreement,
within the quoted uncertainties, with the values by \cite{DOrazi17}  
and
by \citet{Kuzuhara13},
while our stellar structure models do not show a star of solar age as supposed by \cite{Fuhrmann15} and \cite{Bonnefoy18}.

Trying to predict the observed pulsational scenario of GJ~504, 
we used
the ADIPLS package \citep{CD08b} to compute theoretical 
adiabatic oscillation frequencies for all the structure
models satisfying the spectroscopic constraints.
The theoretical result show that the oscillation modes expected to be visible in this star should
be $l=0, 1,2,3$ pure acoustic  modes with frequencies in the range approximately between $(1500-3500)\, \mu$Hz while 
the theoretical large separation calculated by linear fit over the
asymptotic relation for the radial mode frequencies appear to be $\Delta \nu=(98\pm13)\,\mu$Hz.

Among all the possible computed structure models, we selected four models chosen to best-fit the observed  effective temperature, the metallicity and the gravity (Table \ref{tab1}) and with location in the HR diagram shown by coloured  dots (see Fig. \ref{HR}).

In Table \ref{tab:fit} we give a comprehensive set of physical properties for the four different
models  of GJ~504. 
In particular Model~1 has been chosen to match within $1\sigma$ also the luminosity obtained by the SED technique (see Table \ref{tab1}). 

We expect to be able to distinguish among the different models of this target
by measuring at least the large separation in the observed oscillation spectrum.

\begin{deluxetable}{lcccc}
%\tabletypesize{\scriptsize}
%\tablewidth{2pc}
 \tablecolumns{5}
 \tablecaption{Main 
 parameters  for four best-fit structure models of GJ~504.}
 \tablehead{\colhead{} &\colhead{ Model 1}& \colhead{Model 2} & \colhead{Model 3} &\colhead{Model 4}}
\startdata
% 103,2786,
 $M/{\mathrm M}_{\odot}$  & 1.23 & 1.23&1.25& 1.30 \\
                Age (Gyr) & 0.66&  2.46&2.19&  0.74 \\
 $T_{\mathrm{eff}}$ (K)   &6212& 6200 &  6200 &  6205 \\
 $\log g$ (dex)          &4.34&4.22 & 4.25 & 4.32    \\
 $R/{\mathrm R}_{\odot}$ &1.23 &1.42&1.38 &  1.30 \\
 $L/{\mathrm L}_{\odot}  $&2.04&2.69&2.57 &  2.25 \\
 $Z_s$                   &0.026 &0.026&0.027 &  0.028 \\
$X_s$                    & 0.68& 0.67&0.68 & 0.7 \\
$X_c$ &                  0.58& 0.29&0.35 & 0.59\\
$[\mathrm{Fe/H}]$       &0.19 &0.20 &0.21  & 0.21\\
$r_{cz}/R$            &0.840 &0.837&0.846 & 0.846  \\ 
$\alpha_{MLT}$          &1.8 &1.8&1.8 &1.8\\
$\Delta \nu\, (\mu$Hz)&109.6 &87.8& 90.9& 104.5
 \enddata
\tablecomments{$M/{\mathrm M}_{\odot}$ is the mass of the star, $T_{\mathrm{eff}}$ is the effective temperature,
 $\log g$ is the surface gravity, $R/{\mathrm R}_{\odot}$ is the surface radius, $L/{\mathrm L}_{\odot}$ is the luminosity, $Z_s$ is  the surface heavy-element abundance, $X_s$ is the surface hydrogen abundance, $X_c$ is the hydrogen abundance in the core, $[\mathrm{Fe/H}]$ is the iron abundance, $r_{cz}$ is the location of the base of the convective region, $\alpha_{MLT}$ is the mixing-length parameter and $\Delta \nu$ is the large separation obtained from the theoretical pulsational frequencies.}
\label{tab:fit}
 \end{deluxetable}

\section{Observations and data preparation}
\label{sect:obs}
GJ\,504 was observed by TESS during 27 consecutive days of sector 23 from March 18, 2020 to April 16, 2020  with a 2-minute cadence mode.
 During the referee process of this paper, TESS sector 50 observations from March 26, 2022 to April 22, 2022 were available including 20-second cadence data for this star. As explained later in the paper, we also analyzed these data, but the conclusions of the article remain the same. Thus, we describe all the analysis done for sector 23 with 120-s cadence data and we only comment the new sector 50 results when relevant.

In order to perform the seismic analysis and due to the high-level of noise of the TESS 
%120-s cadence 
data for this star, we adopted four different strategies to obtain seismically optimized light curves. In such way we ensure that the obtained results are independent of the methodology applied.

\begin{figure*}[ht!]
    \centering
   \includegraphics[width=12cm]{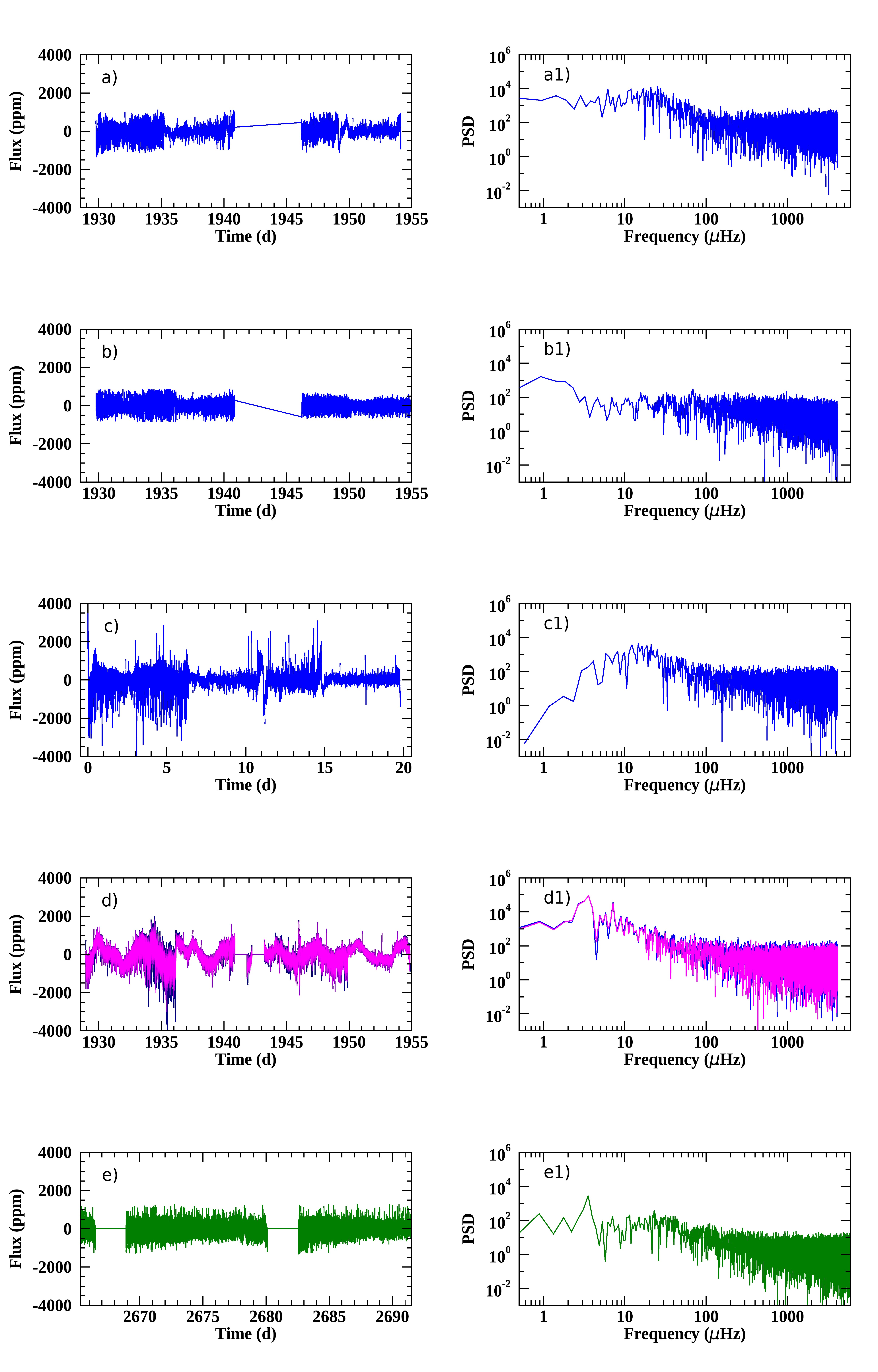}
    \caption{Seismically optimized lightcurves (left figures) and corresponding PSDs (right figures) from the analysis of the 120-s cadence data as explained in Section~\ref{sect:obs}.  The anomalous high scatter observed in this sector is related to the pointing jitter \citep{TESSDataReleaseNotesSector23}.  Panels (e)
depict the results of the 20-s cadence data with the same method applied to get the magenta curves of panels (d).} 
\label{fig:Lcflux}
\end{figure*}

%\subsection{Analysis of 120 second cadence mode data}
The first methodology exploits TESS Science Processing Operations Center \citep[SPOC,][]{2016SPIE.9913E..3EJ} pipeline light curve, with a cadence of 120 s, available on the MAST archive\footnote{\url{https://archive.stsci.edu/hlsp/tess-spoc}}. This raw light curve shows strong modulations at low frequency that is filtered out by applying a smoothing removal process iterated three times. The resultant residuals are subsequently   $3-\sigma$-clipped to eliminate any outliers as depicted in panel (a) of Fig.~\ref{fig:Lcflux}.

The second methodology started with the TESS SPOC 120-s cadence target pixel files. We then extracted a time series for each pixel, rejecting cadences with nonzero quality flags (see for details the TESS Science Data Products Document\footnote{\url{https://archive.stsci.edu/missions/tess/doc/EXP-TESS-ARC-ICD-TM-0014.pdf}}), and constructed an aperture mask using the procedure described in \citet{Buzasi2016} and \citet{Nielsen2020}. Essentially this process produces a time series with the minimum sum of first differences between successive points. We then adopted sigma-clipping at the $4\sigma$ level combined with simple gap filling through the use of a piecewise cubic hermite interpolating polynomial (PCHIP; as implemented in \texttt{Scipy}, \citealt{Scipy}). The result is shown in panel (b) of Fig.~\ref{fig:Lcflux}.

The third approach is based on a filtering of the SPOC lightcurve using two successive Gaussian filters of width 0.25 and 0.125 days. These are 1D convolutional filters, as defined by the Python function \texttt{scipy.ndimage.gaussian\_filter()} \citep{Scipy}. This function uses a Gaussian Kernel that is convolved with the spectrum. The filters are therefore applied to the ensemble of the data. This was done with the aim to remove long periodicities and to reduce the noise level. The lightcurve presents a large gap that could degrade the quality of the spectrum (window effect). The gap was removed before filtering the lightcurve by stitching the second segment to the first one separated by one cadence. Finally, the first time stamp was set to zero. The result is shown in panel (c) of Fig.~\ref{fig:Lcflux}.

\begin{figure}[ht!]
    \centering
   \includegraphics[width=8cm]{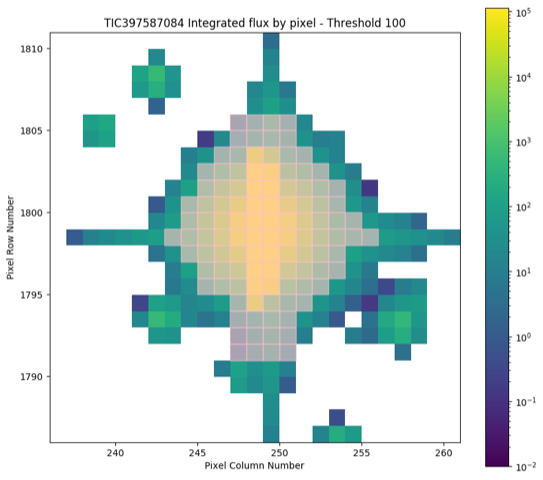}
    \caption{Enlarged mask used in the fourth calibration method described in Sect.~\ref{sect:obs}. The selected pixels are depicted in gray.} 
\label{fig:aperture}
\end{figure}

The last method started from the target pixel file to create a larger aperture. In general, light curves obtained from big apertures are more stable to small instrumental perturbations such as the loss of pointing of the satellite or to the movement of the star during the observations.
To build this larger aperture, contiguous pixels starting from the center of the target are selected. A new pixel is selected only if the integrated flux of the pixel has a negative gradient compared to the previous one (decreasing the flux from the center to avoid any polluting star) and with an average flux greater than a given threshold that has been established to 100 e-/s. Once this is done, an extra pixel at the top and the bottom of the aperture is added to the 4 central rows, which contain several saturated pixels. By selecting these extra pixels, the resulting light curve has smaller dispersion around the mean between the days 1933.5 and 1936. The final aperture is shown in Fig.~\ref{fig:aperture}. It is important to notice that no significant changes were found by adding more pixels to the central rows or by slightly changing the limit threshold of 100 e-/s.

To increase the duty cycle, instead of removing all points with a flag different to zero, we applied two different selections of the NASA quality flags. We either kept all the points except the ones with a flag between 2 and 32 (Dark blue curve in panel (d) of Fig.~\ref{fig:Lcflux}) or
between 2 and 512 (magenta curve in panel (d) of the same figure). We then calibrated the two resulting lightcurves following \citet{2011MNRAS.414L...6G}, 
removing outliers, correcting jumps and drifts. To convert the flux in parts per million (ppm) and remove the low-frequency contribution we used a triangular smooth with a window of half a day. Except for the big gaps in the middle of the run, all the rest were interpolated using inpainting techniques with a multi-scale discrete cosine transform \citep{2014A&A...568A..10G,2015A&A...574A..18P}. These two lightcurves are longer and with some more data in the middle of the run than the other three presented before. 

\begin{figure}[ht!]
    \centering
   \includegraphics[width=8cm]{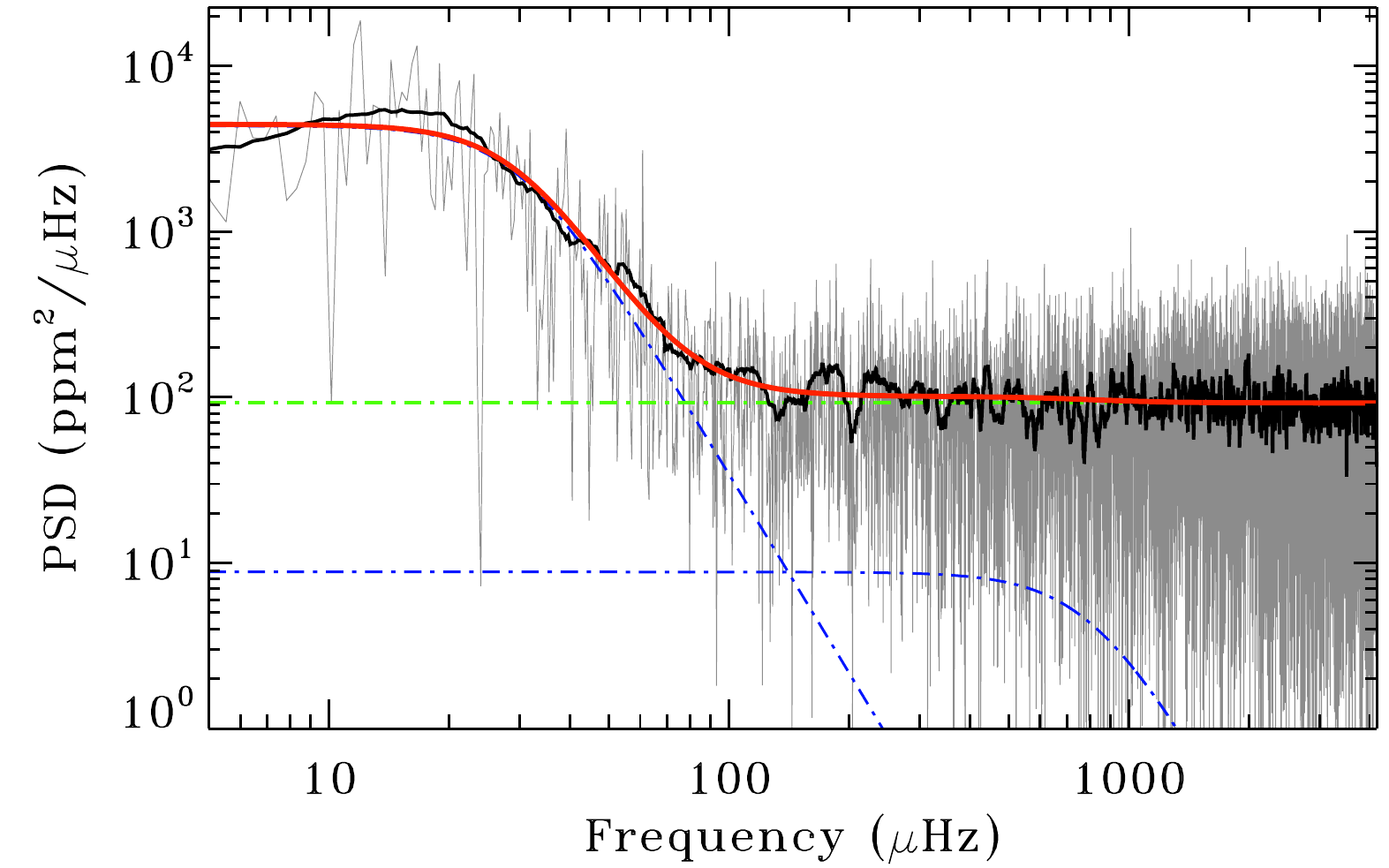}
    \caption{Power spectral density (light gray) with a smoothing overlaid (black curve) of the first seismically optimized lightcurve (shown in panel a) of Fig.~\ref{fig:Lcflux}). The background fit is composed of two Harvey-like profiles (blue curves) and a flat noise level (green curve). The sum of the three components is indicated by the thick red line. No evidence of a Gaussian power excess is found.}
\label{fig:Back}
\end{figure}

%\subsection{Analysis of 20 second cadence mode data}

%The improved precision, according to our calculations, should have allowed us to detect the oscillations and to better constrain the parameters of this star and of its companion by means of asteroseismic techniques.
% In fact, observations with a 20-second cadence might  reduce the level of noise as shown for instance by \citet{Huber2022} and therefore increase the chance of detecting the oscillation power excess. 
 
 The 20-s cadence data observed during sector 50 were also analyzed by the different methodologies described above.
One of the results is
shown in the panels (e) of Fig 3. In this case, the corrections
are the same of the ones applied to produce the magenta curve shown in panel (d), but with a more stringent
high-pass filter with a cut at 0.5 days.

In preparation of the seismic analysis, we computed the Power Spectral Density (PSD). As shown in Fig.~\ref{fig:Back}, the PSD is dominated by flat noise above $\sim$200 $\mu$Hz and a low-frequency slope below $\sim$ 60 $\mu$Hz. Hence, the background can be characterized  by two Harvey components \citep{harvey1985} and a flat noise level. As expected, the high-frequency Harvey profile (related to convective noise) has an amplitude of around an order of magnitude smaller than the flat noise component. The low-frequency Harvey profile, with a knee at around 20 $\mu$Hz, is probably related to magnetism and not convection. 

GJ~504 is indeed a magnetically active star that was part of a large observational campaign, the HK Project, conducted at the Mount Wilson Observatory (MWO) from 1966 to 1995 with the aim to search for stellar analogs to the solar cycle by studying stellar chromospheric activity and variability \citep[]{Wilson68, Wilson78}. These measurements, available from the National Solar Observatory (NSO) website \footnote{\url{https://nso.edu/data/historical-data/mount-wilson-observatory-hk-project/}}, are expressed in term of the dimensionless S-index, defined as the ratio of emission in the Ca II H \& K line cores to that in two nearby continuum reference bandpasses \citep[for further details see, e.g.,][]{Vaughan78, Egeland17}. For this star, within the MWO dataset, about 1342 single measurements are provided in the time interval 1966-1995, allowing us to study its magnetic activity over a time period of nearly 30 years.
%----------------------------------------------------------------------------
\section{Rotation and Magnetic activity analysis}

\subsection{Rotation}
\label{rotation_sec}

\begin{figure*}[ht!]
    \centering
   \includegraphics[width=13cm]{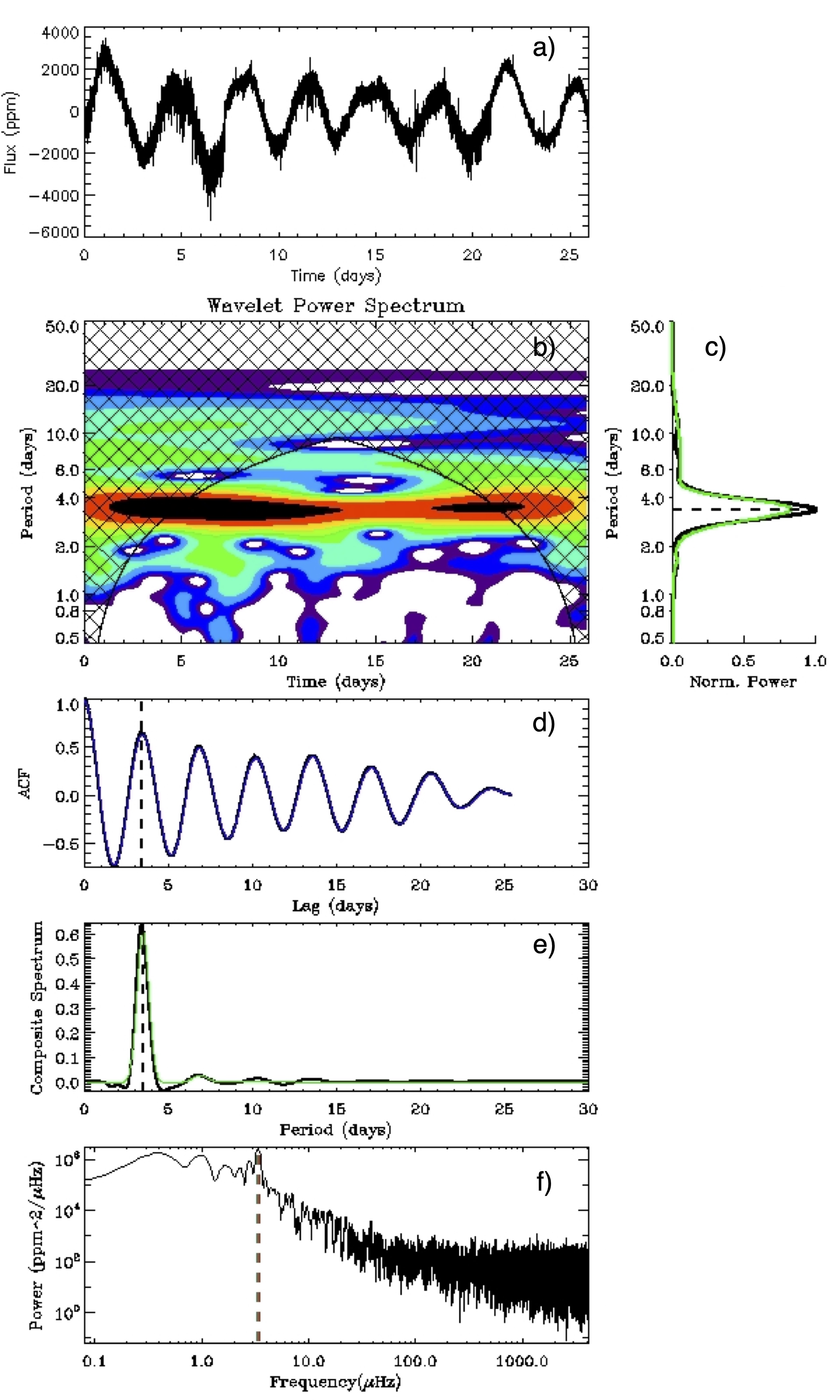}
    \caption{(a) TESS lightcurve for the rotation analysis. (b) Time-period analysis using wavelets. Black corresponds to high power and blue to low power. Black hatched area represents the region that cannot be sampled with the current length of the lightcurve. (c) Projection of the period-time analysis onto the period axis (black) and corresponding fits with multiple Gaussian functions (green). (d) Autocorrelation function. (e) Composite Spectrum (black) and best Gaussian fit (green). (f) PSD in logarithmic scale. The black dotted line indicates the rotation-period estimate.}
\label{fig:Rot}
\end{figure*}

To determine the surface rotation period of GJ~504, a similar methodology as the one applied to the two last lightcurves described in the previous section is employed but this time smoothing the light curve using a triangular filter (double boxcar function). The width of each boxcar is a fifth of the total length. The obtained rotation period, $P_{\rm rot}$, is independent of the flags removed in the lightcurve because we are interested on the long periods and thus the extra removed peaks with a bad flag do not affect the calculation. To look for $P_{\rm rot}$, a methodology combining 3 different techniques is used following, e.g., \citet{2019ApJS..244...21S,2021ApJS..255...17S}. The first method performs a time-frequency analysis using a Morlet wavelet \citep{1998BAMS...79...61T}. The second utilizes an auto-correlation function \citep[e.g.][]{2014A&A...572A..34G,2014ApJS..211...24M}. The third method combines the first two to compute the Composite Spectrum \citep[e.g.][]{2016MNRAS.456..119C}. Hence, a modulation with a periodicity of $(3.4\,\pm 0.25$)\,days is found in the light curve (see Fig.~\ref{fig:Rot}).

We also examined the Mount Wilson data to search for a potential rotational periodicity. The data consist of 1342 observations taken between March 1966 and June 1995. Before searching for the presence of a periodicity in the S-index of GJ 504, we visually analyzed the available Mount Wilson data. We noted that 3 measurements taken in 1993, during the same night, are completely outside the mean range of variation (the average MWO S-index for this star is 0.313), with values that are approximately twice as large. We suspected that such scattered measurements could have arisen as result of an error in the data collection on that night, hence we chose to discard them in the following data analysis of the S-index. We then removed the three outliers with S-index values greater than $4 \sigma$ above the mean, applied a simple linear detrending to the data to remove the lowest-frequency signal, and analyzed the resulting time series using both a DFT and a Lomb-Scargle periodogram. Neither approach results in any significant signal in the range anticipated for rotation, and this conclusion is robust to the inclusion of the 3 deleted measurements. Longer high-quality time series would be necessary to reach any conclusion about the actual rotation period of this star. These results are confirmed with the light curve obtained during the observations of sector 50.

\subsection{Magnetic activity and cycles}
\label{sec:activity_cycle}
While determining the magnetic activity level and the eventually presence of a periodic variability of a star (i.e., a stellar cycle), a key role is played by long-term datasets providing measurements of chromospheric proxies. As known from literature, many stars other than the Sun show a chromospheric variability related to magnetic activity which exhibit periodic variations \citep[see e.g.,][]{Baliunas95, Hall08}. A periodic variability is typically visible also in the photospheric emission, whose phase difference with the chromospheric one reveals the activity dominant regime of the star, i.e. faculae-dominated (phase) or spot-dominated (anti-phase) \citep[]{Radick98, Reinhold19}. 
Before searching for the presence of a periodicity  and to assess its level of magnetic activity, as in Sec. \ref{rotation_sec},
we chose to discard the 3 outliers measurements in the following data analysis of the S-index.

\begin{figure}
	\centering
	\includegraphics[scale=0.48]{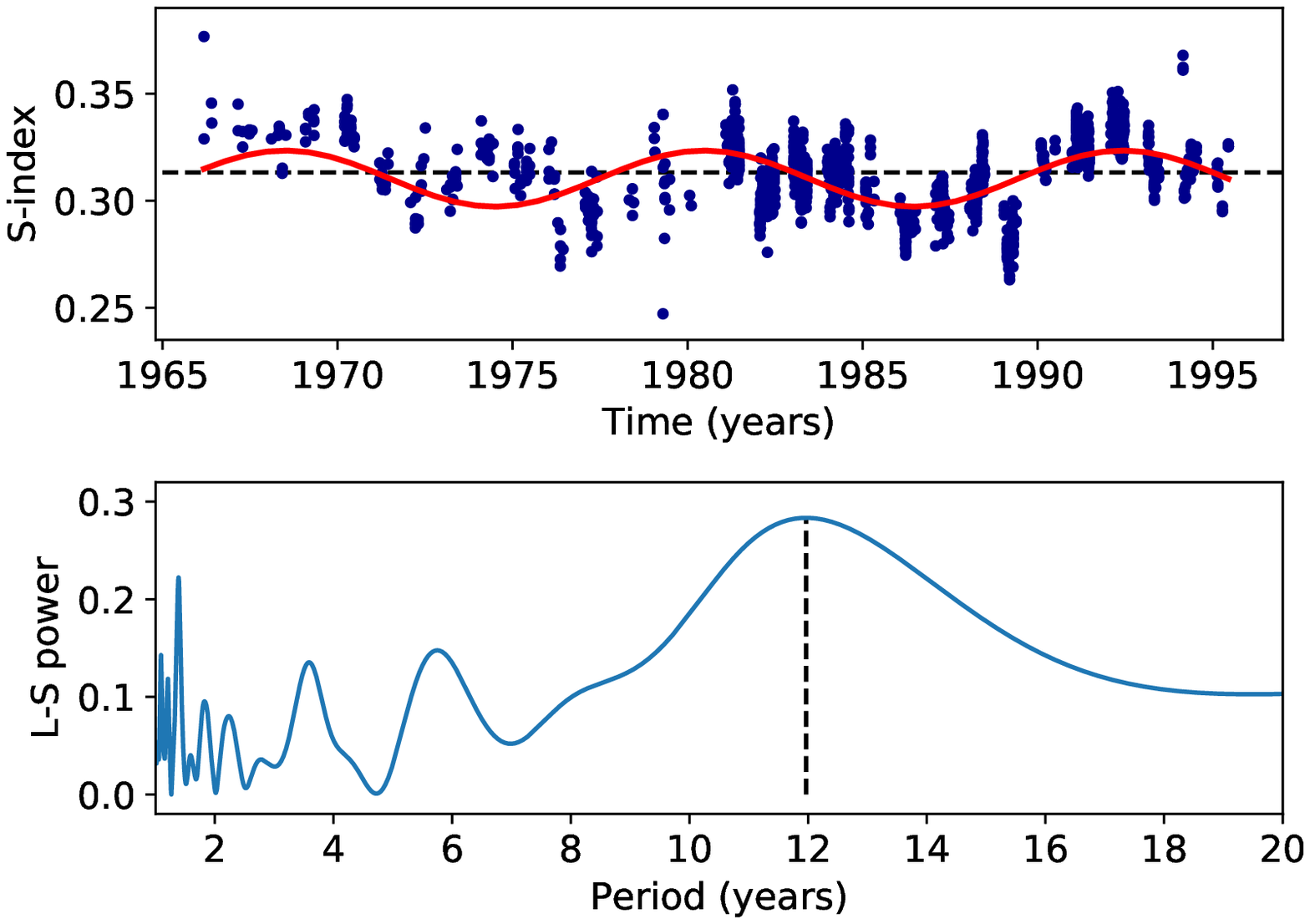}
	\caption{\textit{Top panel:} Mount Wilson S-index measurements for the time interval 1966-1995. The dashed black line indicates the mean S-index. The red line shows the sinusoidal fit to the highest peak as provided by the periodogram in the bottom panel. \textit{Bottom panel:} Lomb-Scargle periodogram of the S-index. The highest peak corresponds to a cycle period of $\simeq 12\, a$.}
	\label{periodogram}
\end{figure}

In order to evaluate the magnetic activity level of the star, we firstly computed the average S-index over the whole time interval (1966-1995) as well as its extreme values. The data are shown in the top panel of Fig.~\ref{periodogram}. The mean S-index is 0.313, while the minimum and maximum values are respectively 0.247 and 0.377. If we compare the mean value with the solar one for cycle 23 (0.170), as reported by \citet{Egeland17}, we can infer that the mean MWO S-index of GJ~504 is around 1.8 times that of the Sun. In addition, the variability in the S-index ($\sim$ 0.13), i.e., the difference between the maximum and minimum values, is greater than that of the Sun during a solar cycle ($\sim$ 0.02). We are, therefore, facing a star whose chromospheric activity level is much higher with respect to that of a reference star like the Sun, pointing towards a probable age smaller than the solar one due to the fact that chromospheric activity typically decreases as the star evolves \citep[]{Skumanich72, Mamajek08, Fabbian17, Gondoin18}.

To search for a long-term periodic variation, the observations of the S-index from the Mount Wilson Observatory available for a large number of stars, constitute a very useful tool. To do that we use an algorithm largely employed in astrophysics, the Lomb-Scargle periodogram \citep[]{Lomb76, Scargle82} which, unlike the more classical Fast Fourier Transform (FFT) analysis, allows to identify periodicity in unevenly sampled data, as in the case of the Mount Wilson observations. The computed Lomb-Scargle periodogram of GJ~ 504 is shown in the bottom panel of Fig.~\ref{periodogram}. Despite the presence of some peaks at small time scales (1.39, 3.59, and 5.75~a), partially due to the data sampling, the highest one corresponds to a main periodicity of 11.97~a. The corresponding false alarm probability (FAP) is 7.97 $10^{-93}$, indicating that the detected cycle period is statistically significant and unambiguous. This result indicates for this star the presence of a principal periodic chromospheric variability with a characteristic time quite similar to the Sun Schwabe 11-year cycle, even if it is set to a higher level of activity compared to the latter.

In addition to measuring the magnetic activity of GJ~504 with spectroscopic data, we also computed the photometric magnetic activity index, $S_{\rm ph}$ using TESS data. Following \citet{2014JSWSC...4A..15M,2014A&A...562A.124M}, it is computed as the standard deviation of subseries of length 5\,$\times P_{\rm rot}$ to ensure that we are measuring the variability due to the magnetic activity. From that temporal $S_{\rm ph}(t)$, we take the mean value. Using the rotation period of 3.4 days found in Section~4.1, we obtain $S_{\rm ph}= (1231 \pm 7.8)$\,ppm.

%---------------------------------------------------------------------------
\section{Searching for solar-like oscillations}
\label{sec.osc}
Based on the spectroscopic parameters of GJ\,504, we looked for the solar-like oscillations using the prediction from Sects~\ref{Sect:pred} and \ref{models}. As seen on Figure~\ref{fig:Back}, the modes are not obvious and hence, we applied global seismic methods to look for the global seismic parameters.

In order to confirm the results obtained, the analysis of the power spectrum was independently performed by 4 teams who adopted different methods as described below.

The first method consisted in searching for the presence of oscillations in a region centered around $2000~ \mu$Hz, as suggested by the former predictions for $\nu_\mathrm{max}$. For this purpose we adopted the public tool DIAMONDS\footnote{https://github.com/EnricoCorsaro/DIAMONDS} \citep{Corsaro14} coupled with the Background code extension\footnote{https://github.com/EnricoCorsaro/Background} for estimating the level of the background signal. The background signal, as described in \cite{Corsaro17b}, comprises two Harvey-like profiles accounting for possible granulation-related signal and other variations at low frequency, a flat instrumental noise, and a Gaussian envelope of the solar-like oscillations. In particular we performed a Bayesian model comparison by means of the Bayesian evidence computed by DIAMONDS to select the best competing background model between one including the Gaussian envelope of the solar-like oscillations and one excluding it (see also \citealt{Mullner21}). The resulting Bayes' factor suggests that the incorporation of an additional Gaussian profile is not statistically justified, meaning that in the light of the current data-set we could not detect the presence of a power excess due to stellar oscillations in this star. This result is depicted in Fig.~\ref{fig:Back}, where no clear power excess due to stellar oscillations can be observed.

With the second method (applied to the second set of lightcurves), we searched for local peaks in the amplitude spectrum, requiring a minimum separation between peaks of $20~\mu \rm Hz$, and generated an upper envelope across those peaks using cubic spline interpolation. The location of the maximum of that envelope was taken to approximate $\nu_{max}$, and was estimated by fitting a simple (linear + Gaussian) model and taking the center of the Gaussian to represent the location of the maximum. Uncertainties were estimated by repeating the procedure 1000 times with the minimum separation between peaks allowed to vary randomly between [0, 40] $\mu \rm Hz$. In each case,  we also tried to estimate the large separation by performing an autocorrelation of the central $400~\mu \rm Hz$ of the amplitude spectrum.  Unfortunately this process resulted in a peak envelope height not large enough to claim for a statistical significance detection.
%This process resulted in $\nu_{\rm max} = 1919 \pm 40 ~\mu \rm Hz$ and $\Delta \nu = 85.0 \pm 3.6 ~\mu \rm Hz$. However, the peak envelope height is not large enough to claim statistical significance for this result, though it is suggestive. 
%and we will use this value for comparison with the theoretical models.

The third pulsation search algorithm is an upgraded version of \cite{Benomar2012} and was applied to the third lightcurve described in Section~\ref{sect:obs}. The first step consists in getting initial guesses for a Bayesian analysis that follows if pulsations are detected conclusively. 
A first power spectrum F$_{noise}$ of the star is produced by heavily smoothing (box-car smoothing of width $\approx 100$ $\mu$Hz) the original power spectrum. This allows to have an approximation of the noise background as pulsations (if any) are damped by the smoothing.
A second spectrum F$_{modes}$ is produced using a smoothing coefficient (box-car smoothing of $\approx 0.8$ $\mu$Hz) optimised for revealing individual pulsations.  
The maximum of amplitude of the ratio F$_{modes}$/F$_{noise}$ is then estimated by performing a local $3rd$ order polynomial fit. The FWHM of the polynomial curve is used to have a first estimate of the potential region for pulsations and the Height-to-Noise ratio is used to evaluate the significance. We found only a marginal detection of pulsation.
%at  $\nu = 2040 \pm 282$ $\mu$Hz. 
To confirm the detection, a fit of the power spectrum is performed. It involves describing pulsations with a gaussian envelope and the noise background with two Harvey-like profiles \citep{harvey1985} and white noise. Unfortunately, the Bayesian Maximum a Posteriori estimates gave us a significance for pulsations below $1\%$ when compared to a pure noise fit of the spectrum.

Another team analyzed two sets of lightcurves (LC1 and the one with our own aperture) with the A2Z pipeline \citep{2010A&A...511A..46M}. Briefly, they looked for the mean large frequency spacing by computing the power spectrum of the power spectrum. We then fitted the background with three components: a Harvey law to model the granulation where the slope was fixed to 4, a Gaussian function for the modes and the white noise. After subtracting the background without the Gaussian function, we fitted another Gaussian function to estimate the frequency of the maximum power.  A blind run of the A2Z pipeline found some excess of power around 1000\,$\mu$Hz, but no frequency spacing that agrees with the global seismic scaling relations \citep{Kjeldsen95} was measured with a high level of confidence level. By forcing the pipeline to look around 2000\,$\mu$Hz, 
%a frequency spacing of $\sim$84\,$\mu$Hz is measured with more than 95\% confidence level but 
no Gaussian fit converged to obtain $\nu_{\rm max}$. 
These results lead to a non detection of the modes with the A2Z pipeline.
 
Finally we also computed the Enveloppe Auto-Correlation Function (EACF) following \citet{Mosser2009}. No detection of $\Delta \nu$ was done with this method. This is not surprising as the EACF and A2Z methods give similar results as shown for a sample of low signal-to-noise ratio targets where both methods were used \citep{Mathur2022}.

 The analysis of the ultra short 20-s cadence data obtained during the TESS observations  in sector 50 did not provide any clearer conclusion.
 %The new PSD is still dominated by noise and we were unable to obtain a statistically significant detection.
 We estimated the white noise level in the 20-s cadence data to be about 1.7 ppm$^2/\mu$Hz, which is almost one order of magnitude (9.6 times) smaller than the one measured in the 120-s cadence data obtained from sector 23 (were we find 16 ppm$^2/\mu$Hz instead). Despite this notable improvement in the level of noise, we were unable to obtain any statistically significant detection of a power excess due to solar-like oscillations.
 Moreover, we stitched together the data of both sectors to have a longer light curve. To do so, both light curves were filtered with a high-pass triangular filter with a cut at 0.5 days and sector 50 data were re-binned to 120-s. To remove the long gap between the two sectors, the time of the first point of sector 50 was changed to 120 seconds after the last measure of sector 23. This has no influence on the p modes as they are expected to have shorter lifetimes than two years. Once again, we were not able to detect any excess of power.
 %only a hint of modes were found around ~2 mHz.
% {\bf Finally, we also computed the auto-correlation function of a smoother PSD (over 50points) in the frequency range 1700 to 2200\,$\mu$Hz, where the modes are expected. No clear pattern showing the different separations of the modes (small separation, mean large frequency separation) is obtained as already done for other solar-like stars (REFS: e.g. Mosser and Appourchaux 2009, Chontos et al. 2021). The highest peak is found at 34.2\,$\mu$Hz but does not fit any of the scaling relations even assuming that we measure half of $\Delta \nu$. Furthermore, the peak has a small amplitude, leading once again to a non-detection of the modes.}

%---------------------------------------------------------------------------
\section{Discussion }

%----------------------------------------------------------------------------

\subsection{Impact of magnetic activity on the solar-like oscillations}
 We discuss here the analysis of the TESS data  and the non-detection of pulsation modes on the solar-like star GJ~504.  While some of the analysis pointed toward a possible excess of power in the region around 2000~$\mu$Hz, no reliable detection of solar-like oscillations can finally be reported. It is possible that this might be due to the high noise of the TESS data.
 However, another possible explanation can be attributed to the presence
of a high level of magnetic activity. In fact, several authors have already shown that magnetic activity is responsible for suppression of solar-like oscillations as already found in several targets \citep[e.g.][]{Garcia2010,2011ApJ...732L...5C, Mathur2019}.
The evolutionary stage, the estimate of the age and the analysis of the magnetic activity indices of GJ~504 as developed in Sects. \ref{sec:activity_cycle} reveal
a level of magnetic activity typical of young solar-like objects \citep[e.g.][]{Bohm2007,Hall2007}.
In fact, the analysis of the chromospheric emission, through the S-index, has highlighted a fairly high level of magnetic activity (mean S-index = 0.313), $\sim$ 1.8 times that of the Sun. The study of the periodicities with the Lomb-Scargle algorithm has pointed out a main principal cycle at  11.97 a, in agreement with the $11.79\pm0.28$\,a detected cycle by \cite{BoroSaikia2018}, but also revealed the presence of other smaller amplitude cycles. The coexistence of different cycles is a typical characteristic of fast rotating stars, where a higher number of dynamo modes are excited \citep[]{Durney1981, Olah2016}, as it is the case of this star for which we found $ P_{\rm rot} \simeq 3.4$ days. 

Once the stellar rotation and the main activity cycle period are known, we can compute the ratio $P_{\rm cyc} / P_{\rm rot}$, a quantity which is known to be related to the dynamo number $N_D$ \citep[see e.g.,][]{Soon1993, Baliunas96}. For stars older than 2.5 Gyrs, like the Sun, the quantity log$(P_{\rm cyc}/P_{\rm rot})$ is typically around 2 (see Fig. 6 in \citealt{Olah2016}), while we obtain  3.1. This result indicates that GJ~504 is an active star with an age smaller than the one where the transition from spot to faculae domination, associated with a Rossby number $ \mathrm{R_0}\sim$1 and an age $\sim$ 2.55 Gyr \citep{Reinhold19}, is believed to happen. This is in agreement with our age estimation of $1.3\pm1.3$ Gyr. This is also consistent with the fact that \cite{Reinhold19} found the photometric and chromospheric variability to be out of phase ($\Delta\phi = 0.34$), indicating that the star is still in the spot-dominated activity regime which characterizes the young and active stars.

 Concerning our attempt to detect solar-like pulsations, we used the calibrated formula by \cite{Bonanno14}, to relate the Mount-Wilson chromospheric $S$-index to the global oscillation amplitude $A_\mathrm{max}$. By using the mean $S$-index obtained in Sect.~\ref{sec:activity_cycle}, we obtain for this star a regime of significant oscillation amplitude suppression (see, e.g., Fig. 2 of \citealt{Bonanno14}), defined by an expected global oscillation amplitude of $A_\mathrm{max} = 1.6$\,ppm, which is rather low as compared to the level of background found in the data. This magnetic activity suppression likely justifies the non-detection of an oscillation power excess in the stellar power spectrum. Even in the case of minimum of activity, corresponding to an $S$-index of 0.247 (see Sec. \ref{sec:activity_cycle}), the expected oscillation amplitude would be $2.6$\,ppm, which is lower than the average background noise measured in the TESS data estimated to be 6.3\,ppm from the background fitting we performed.

 In addition, as obtained in Sect.~\ref{sec:activity_cycle}, the $S_{\rm ph}$ of this star is $(1231\pm 7.8)  {\rm ppm}$ during the TESS observations. Knowing that for the Sun, the average $S_{\rm ph}$ value is 166.1\,ppm, we must conclude once more that GJ~504 appears to be very active (7.5 times higher than the Sun) in agreement with the result obtained from the spectroscopic observations. Comparing this level of activity with the stars with and without detection of modes \citep[see Figure 10 of][]{Mathur2019}, only 3 stars with a detection of solar-like oscillations have an $S_{\rm ph}$ above 1000\,ppm. For these stars, the comparison of the amplitude of the modes observed in the {\it Kepler} data and the predicted amplitude gives that $A_{\rm max,obs}/A_{\rm max,pred}\sim0.8$ in average, varying between 0.70 and 0.93. This means that we can have a reduction from  7 to 30\% in the amplitude of the modes. 
 In the case that we are dominated by the noise, this can even add up to the explanation of the non detection of the modes in this star.  Note that these three {\it Kepler} stars are metal poor (with [Fe/H] around -0.2dex), which according to \citet{2011LNP...832..305S} can lead to higher amplitudes and could counter-balance the effect of the surface magnetic activity.

\section{Conclusion}
\begin{deluxetable}{lc}
%\tabletypesize{\scriptsize}
%\tablewidth{2pc}
\caption{The parameters of GJ~504 as derived by the present analysis based on the use of TESS and Mount Wilson data.}
 \tablecolumns{2}
 \tablehead{\colhead{} &\colhead{Present value}}
 \startdata
 $P_{\rm cyc}$ (a) & $11.97$ \\
$P_{\rm rot}$ (d) & $ 3.4\pm0.25$ \\
 $S$-index  & $0.313\pm0.07$ \\
 $S_{\rm ph}$ (ppm)  & $1231\pm7.8$ \\
$Age$  (Gyr) & $ 1.3\pm1.3$ \\
$M/{\mathrm M_{\odot}}$ & $1.28\pm0.07$\\
$ R/{\mathrm R_{\odot}} $ & $1.38\pm0.20$\\
$\Delta \nu$ ($\mu$Hz) & $ 98\pm13$\\
%$\nu_{\rm max}$ ($\mu$Hz) & $1919\pm 40 $
 \enddata
\tablecomments{$P_{\rm cyc}$ and $P_{\rm rot}$ are  the magnetic activity cycle and the
surface rotation period respectively. $M/{\mathrm M_{\odot}}$ is the mass of the star, $R/{\mathrm R_{\odot}}$ is the surface stellar radius, while $\Delta \nu$ is the theoretical large separation value as obtained by stellar modelling.
%, while  $\Delta \nu$ and  $\nu_{\rm max}$ are the dubious values of large separation and frequency of maximum amplitude of oscillation provided by the present analysis 
.}
\label{tab:results}
 \end{deluxetable}
In this article we present a new attempt to study the solar-like star GJ~504, observed by the space mission TESS and known to host an exoplanet with values of mass and radius not yet confirmed.
Unfortunately, we did not succeed to characterize this star by means of asteroseismic techniques, since we did not find evidence for a clear excess of power, 
% although our pipelines were able to find a marginal signal  of periodicity around $2000-2500~\mu$Hz that could be related to the large separation, but not significant enough to be claimed as presence of solar-like oscillations.

With the aim to reach a substantial step forward in the characterization of this star by clearly detecting the solar-like oscillations, we proposed through the Director discretion time (DDT) to observe it again during sector 50 with 20-second cadence mode.
 Based on TESS magnitude of $4.6552 \pm 0.0073$ \citep{stassun2019},  the analysis of the 20-second cadence data should have yield an improvement in photometric precision of $\simeq30\%$ due to the reduced influence of pointing jitter on cosmic-ray rejection for bright stars \citep{Huber2022}.  Furthermore, the measured period $P_{cyc}\simeq 12$~a of the main magnetic cycle of GJ 504 implies that the stellar magnetic cycle minimum should occur between 2022 and 2023, perhaps overlapping with TESS Sector 50, leading to the best conditions to minimize the amplitude-suppressing effect of magnetic activity. Unfortunately the analysis of the more recent data did not allow the hoped detection.

%although a marginal detection of pulsations seems to manifest at about $\nu_{\rm max}=(1919\pm40)~\mu$Hz with some indication of large separation with $\Delta \nu=(85\pm3.6)~\mu$Hz. 
This non detection can be explained by the high level of magnetic activity for the star. Indeed the spectroscopic analysis yields an S-index of 0.313 and the photometric analysis of the TESS light curves provides a magnetic proxy $S_{\rm ph}$ of 1231\,ppm, both indices being much larger than the solar value of 0.170 and 161\,ppm respectively. 
Given the values of the S-index and $S_{\rm ph}$, the modes are predicted to suffer an important decrease of their amplitudes, probably close to the noise level in the TESS observations.

Nevertheless, all the results that we have deduced by analyzing the photometric data by the TESS space mission, supported by the measurements collected by the Mount Wilson Observatory long term campaign spanning nearly 30 years and by the modelling procedures, have allowed us to get important conclusions on the large debated parameters of this target. In Table \ref{tab:results} we summarize the stellar parameters that best represent GJ\,504.

Firstly, the analysis of the three decades long Mount Wilson spectroscopic observations yields the detection of a main magnetic cycle of 11.97\,a and, at least, other two smaller amplitude cycles of 5.75~a and 3.59~a.

Further, the analysis of the Mount Wilson data did not
allowed us to measure a stellar rotational period, while the TESS light curves show a  clear modulation corresponding to $P_{\rm rot}=$3.4\,d.

The stellar radius and mass have been calculated from stellar models constrained on spectroscopic measurements of gravity, metallicity and effective temperature only. 
Moreover, the value of the stellar radius results in agreement with independent measurements obtained in the present article by applying the SED and the SBCR methods,
but also with the more accurate interferometric radius
determined by \citet{Bonnefoy18}.

The age of GJ~504, as obtained by stellar modelling based on accurate spectroscopic fundamental parameters, appears to be $Age\leq2.6$\,Gyr in agreement, within the quoted uncertainties, with previous finding by \cite{DOrazi17} and \cite{Kuzuhara13}. In particular, the rotational period and the main magnetic cycle locate this G-type star in the regime of chromospheric activity dominated by the superposition of several magnetic cycles  during which, as according to the not yet confirmed theory of \cite{VanSaders2016} and \cite{Metcalfe2017}, the
magnetic braking should still be acting while the rotation is slowing down.
This situation puts this target well before the magnetic transition, which would bring this star at the age of about 4-5~Gyr to the shutdown of the magnetic braking reaching a low activity state.

 Adopting this new age value, along with {\sc SPHERE} $JHK_1$K$_2$ photometry for the companion \citep{Bonnefoy18} and the COND-AMES model atmospheres \citep{2003Baraffe}, we gather companion mass and radius of $M_{\rm P}=(16.5\pm4.8){\mathrm M_{Jup}}$ and $R_{\rm P}=(1.00\pm0.03 )\mathrm{R_{Jup}}$ (the related errors are simply the standard deviation from the four photometric bands, so they are certainly under-estimated). 
Hence, given the large uncertainty in age, we cannot confirm/disprove from the present study whether GJ\,504b is located in the brown dwarf or planetary regime.
We confirm that the actual scenario is compatible with the hypothesis of engulfment of a sub-stellar companion, as previously proposed by \cite{Fuhrmann15} and \cite{DOrazi17}, necessary to explain the low rotational period and the age of the star. In fact, the orbits of the planets can change in time due to several mechanisms, such as tidal
interactions, stellar winds, planet evaporation, leading a planet to be engulfed by its host star \citep{Privitera2016, Benbakoura2019}. As a planet moves to inner orbits, conservation of angular momentum of the system imposes that  a reduction in the orbital angular momentum is compensated by the increase in the stellar rotation. \citet{Benbakoura2019} showed that ultra hot Jupiters at closer orbital distances could spin up their hosts during the main sequence, while lighter planets (less than $1M_{jup}$) could not.
A possible signature of a planet engulfment could be, for example, an anomalous metallicity. However, this target does not seems to be overmetallic, as shown in Table \ref{tab1}.

%Based on TESS magnitude of $4.6552 \pm 0.0073$ \citep{stassun2019}, the 20-second cadence data should yield an improvement in photometric precision of $\simeq30\%$ due to the reduced influence of pointing jitter on cosmic-ray rejection for bright stars \citep{Huber2022}.  Furthermore, the measured period $P_{cyc}\simeq 12$~a of the main magnetic cycle of GJ 504 implies that the stellar magnetic minimum should occur between 2022 and 2023, perhaps overlapping with TESS Sector 50, leading to the best conditions to minimize the amplitude-suppressing effect of magnetic activity. 

 Finally, since we believe that GJ~504 might represent an ideal target also for the ESA/PLATO \citep{Rauer2016} space mission, with scheduled launch in the end of 2026, we verified that it is included in the all-sky PLATO input catalogue  \citep{Montalto2021}, with the name of PIC~DR1~35698898.
However, according to the proposed PLATO fields presented in \cite{Nascimbene2022}, the portion of the sky where GJ~504 is located will not be considered for the two first years of the PLATO observations.

%****Acknowledgements****

\begin{acknowledgements}
%The publication of the present article has been supported by INAF through IAPS FFO.
This paper includes data collected with the TESS mission, obtained from the MAST data archive at the Space Telescope Science Institute (STScI). Funding for the TESS mission is provided by the NASA Explorer Program. STScI is operated by the Association of Universities for Research in Astronomy, Inc., under NASA contract NAS 5–26555. R.A.G. Acknowledges funding from the PLATO CNES grant. S.M.\ acknowledges support by the Spanish Ministry of Science and Innovation with the Ramon y Cajal fellowship number RYC-2015-17697 and the grant number PID2019-107187GB-I00. D.B. acknowledges support from the National Aeronautics and Space Administration under the Living With A Star program, grant number NNX16AB76G. R.R. is a PhD student of the PhD course in Astronomy, Astrophysics and Space Science, a joint research program between the University of Rome “Tor Vergata”, the Sapienza University of Rome and the National Institute of Astrophysics (INAF). The authors thank the anonymous reviewer for the valuable help in improving the manuscript.
\end{acknowledgements}

\bibliography{references}

\end{document}